\documentclass[a4paper,11pt]{article}
\usepackage{jheppub} 
\pdfoutput=1
\usepackage{graphicx,array}
\usepackage[dvipsnames]{xcolor}
\usepackage{amstext}
\usepackage{amssymb}
\usepackage{slashed}
\usepackage{cancel}
\usepackage{subcaption}
\usepackage{booktabs}
\usepackage{hyperref}

\usepackage{pifont}
\newcommand{\fig}[1]{fig.~\ref{#1}}%

\definecolor{orange}{rgb}{1,0.5,0}
\definecolor{brown}{rgb}{0.59, 0.29, 0.0}
\definecolor{note_fontcolor}{rgb}{0.80078125, 0.80078125, 0.80078125}

\usepackage{tikz}
\usetikzlibrary{arrows,shapes.misc,decorations.markings,decorations.pathmorphing,positioning,intersections}

\usepackage{feynmp}
\usepackage{graphicx}
\usepackage{slashed}
\usepackage{subcaption}
\usepackage[bottom]{footmisc} 

\newcommand\snowmass{\begin{center}\rule[-0.2in]{\hsize}{0.01in}\\\rule{\hsize}{0.01in}\\
\vskip 0.1in Submitted to the  Proceedings of the US Community Study\\ 
on the Future of Particle Physics (Snowmass 2021)\\ 
\rule{\hsize}{0.01in}\\\rule[+0.2in]{\hsize}{0.01in} \end{center}}

\title{
\begin{center}
Warped Compactifications in Particle Physics, Cosmology and Quantum Gravity
\end{center}
}

\author[a]{Prateek Agrawal}
\author[b]{, Cari Cesarotti}
\author[c]{, Andreas Karch}
\author[b]{, Rashmish K.~Mishra}
\author[b]{, Lisa Randall}
\author[d]{, and Raman Sundrum}

\affiliation[a]{Rudolf Peierls Centre for Theoretical Physics, University of Oxford, Parks Road, Oxford OX1 3PU, UK.}
\affiliation[b]{Harvard University, 17 Oxford Street, Cambridge, MA, 02139, USA.}
\affiliation[c]{Theory Group, Department of Physics, University of Texas, Austin, TX 78712, USA.}
\affiliation[d]{Maryland Center for Fundamental Physics, Department of Physics, University of Maryland, College Park, MD 20742, USA.}

\abstract{Particle physics has evolved in the past decade through evaluating the consequences of experimental measurements as well as exploiting theoretical tools that permit exploration of new model building and cosmological possibilities. Particularly due to insights from the AdS/CFT correspondence, higher-dimensional warped compactifications, in particular, have played a big role in recent developments by allowing a study of regimes of parameters that would otherwise be intractable. Similarly, theoretical developments in quantum gravity benefit from the bigger range of possibilities that can be explored using warped geometry, allowing for constructions of string vacua with positive cosmological constant and for the exploration of entanglement and information transfer in arbitrary dimensions. Puzzles remain in both more phenomenologically oriented and more theoretically oriented contexts which form the basis for a rich research program in the future as well.
}

\begin{document} 
\snowmass
\maketitle
\flushbottom

\section{Introduction}
The idea of extra dimensions---first introduced to physics in the early twentieth century---had a renaissance in the late 1990s when  potential phenomenological applications were recognized and the notorious chirality problem was solved with branes and string constructions. As an outgrowth of that effort, new solutions to Einstein's equations---in particular warped extra-dimensional geometry with localized gravity--motivated in part by string theoretical constructions with branes---were discovered whose implications have had far-reaching consequences well beyond the initial phenomenological domain but including those as well. An important connection and reason for this ubiquity is the relation to AdS spacetime, where many new and exact results have been found. Because of the diverse phenomenological and cosmological applications, the discovery of a truly novel model building element, and the potential to exploit dualities to better understand strong interactions and conceptual aspects of the gravitational theory, warped extra dimensional geometries have had a sustained and consequential impact and are likely to do so in the future.

This white paper focuses on the relevance of warped extra-dimensional geometry to phenomenology, cosmology, gravity wave physics, string constructions of de Sitter space, and the black hole information paradox--all of which have been active research topics over the past decade and which lead to intriguing questions promising exciting developments in the future. The first few topics relate to some initial work on the hierarchy problem, whereas the latter topics re-emerged from  work in  string theory trying to find tractable realizations of de Sitter space and the search for workable models to study black hole entropy and evaporation (see \cite{Almheiri:2020cfm} for a review), notably in more than two dimensions. In this white paper we summarize some of the ways the geometry appears in a variety of contexts that can feature in the theory initiative--particularly in its attempt to unify the cohort of activities currently underway.
\section{Model Building and Phenomenology}

\begin{figure}
    \centering
    \includegraphics{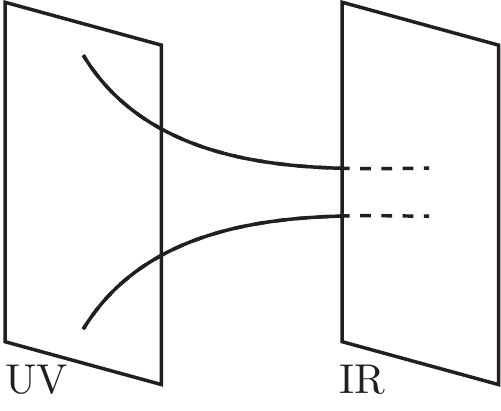}
    \caption{\small{Illustration of a finite extra dimension with warped geometry.}}
    \label{fig:rsModel}
\end{figure}

A major driving force for many model building and  phenomenological activities in high energy physics in the last few decades has been the seemingly unnatural hierarchies in the Standard Model (SM)---the Planck Electroweak hierarchy, and the fermion mass hierarchies (and the associated flavor structure).  The initial work  in the late 90s on Randall-Sundrum (RS) models, based on warped extra-dimensional geometry (illustrated in \fig{fig:rsModel}), was a new Beyond Standard Model (BSM) framework to maintain the EW hierarchy and as an added feature could naturally account for fermion hierarchies, making it one of the few of the electroweak theories to naturally account for both hierarchies with a single underlying mechanism. Such theories address the flavor structure of the SM without pushing the scale for new physics to a very high value, which is marginally acceptable due to the RS GIM mechanism~\cite{Agashe:2004cp,Cacciapaglia:2007fw,Fitzpatrick:2007sa,Perez:2008ee}. The low scale of new physics poses challenges, but also in principle makes the framework  more readily testable than most flavor models--although signals can still be challenging. The experimental efforts to look for new physics in this framework, both by direct production, or by precision measurements, has been one of the drivers of our experimental programs--often in ways applicable to other BSM scenarios.

Moreover, the framework can naturally explain neutrino masses and large mixing angles in the lepton sector alongside small mixing angles in the quark sector. The wide variation in the fermion masses and mixing angles in the SM is one of the biggest challenges for any new physics scenario. Instead of models that address parts of the puzzles separately, it is desirable to have a framework that can give this wide spectrum in a natural way, from a same underlying mechanism. This is particularly challenging, given anarchic neutrino masses but hierarchical quark masses. RS models based on warped extra-dimensions provide a natural way to  generate such hierarchies using anarchic 5D Yukawa structure, using the peculiar dependence of the fermion wavefunction on the bulk fermion mass---exponential in the IR region, but power law in the UV region.  In particular, to explain neutrino masses and large mixing angles in the lepton sector, ref.~\cite{Perez:2008ee} considered Majorana mass for the right-handed neutrinos and a standard type I seesaw mechanism in the warped scenario. The intermediate seesaw scale needed in type I seesaw was naturally realized in the 5D framework without putting in a new scale via a natural choice of 5D mass parameters. 
In a recent work, ref.~\cite{Agashe:2015izu} considered the same setup for generating neutrino masses, but by going to the mass basis rather than the KK basis, they showed that the seesaw mechanism looks inverted. The very small neutrino masses now arise from exchange of singlet modes that are mostly Dirac, but have a highly suppressed Majorana mass term, the latter being a natural consequence of the exponential nature of the wavefunction.

%

Since the flavor aspects are  covered in another white paper~\cite{flavorSnowmass2021}, the focus of our model building section will be recent work on naturalness. To address the lack of experimental discoveries, new research has 
introduced modifications with varying amounts of non-standard signals, so that the present experimental analyses are not directly applicable. This aspect can be taken as the starting point, but the implications are broader, since toy-models in existing BSM frameworks where the resulting signal is challenging can provide a way to develop and therefore cover a wider class of signal sensitivity in our experimental programs.

%

One possible direction along these lines has been  the construction of  UV complete models that can arise within existing BSM frameworks. In one such approach, ref.~\cite{Agashe:2016rle} constructed a 5D model with a non-standard KK spectrum. The 5D bulk gauge and matter fields were required to propagate to different degrees in the IR of the warped geometry, resulting in a qualitatively different low energy effective field theory (EFT) description, and hence drastically different signals at colliders. The resulting KK spectrum where the matter KK modes are heavier than the gauge KK modes, allowed the models to be safe from flavor/CP constraints. The collider accessible modes are the KK graviton and KK gauge bosons, although with very different decay patterns as compared to the minimal RS scenario. An important feature was a relatively lighter radion in the spectrum. The phenomenological consequences of these features was analyzed in more detail in refs.~\cite{Agashe:2016kfr,Agashe:2017wss,Agashe:2018leo}, and 
a full exploration of the resulting low energy EFT and their experimental consequences has been useful from the point of covering the space of possibilities, though sometimes requiring moderate fine-tuning in the EW sector.  
Dedicated experimental analysis based on these scenarios were performed recently~\cite{CMS:2021nrj,CMS:2021qyp,CMS:2021qev}.

In another work with the same broad motivation, ref.~\cite{Csaki:2018kxb} constructed a 5D model where the KK states form a continuum beyond a certain scale--a qualitatively different spectrum of new physics states. The explicit construction used warped geometry in a linear dilaton background. The top and gauge partners needed to regulate the Higgs mass sensitivity form a continuum, though a realistic benchmark still required percent level fine-tuning. The continuum nature of the KK states motivates broader experimental search strategies, given that the standard resonant searches are inapplicable. Both these works explored interesting modifications to the low energy EFT in this framework, and their non-standard experimental signals--these directions are well motivated independently in the search for new physics.

Unnaturalness of the cosmological constant (CC) is an even more challenging puzzle, at least as measured by the degree of unnaturalness. Even though much of past research has focused on them separately, it is plausible that the two are related, and warrants constructing explicit models where such connections can be studied. In this direction, ref.~\cite{Arvanitaki:2016xds} constructed a model based on flat as well as warped extra-dimensions, where tuning the mass of Higgs relative to a fermion mass scale allowed the radius of an extra dimension to become large and develop an enhanced number of vacua available to scan the cosmological constant down to its observed value. The model has fermions that are accessible at colliders, and an unnaturally light radion, which evades high energy collider constraints.
In another explicit model, ref.~\cite{Bloch:2019bvc} constructed a crunching mechanism that selects Hubble patches with small CC after reheating.
Patches with large CC tunnel at a fast rate to a vacuum with large negative CC and ultimately crunch, while patches with small CC are long-lived. A crucial ingredient is a crunching sector that reacts to the presence of a large CC and acts to lower its value, thereby driving the relevant patch to crunch. The crunching sector has a super-cooled first order phase transition that was modelled by warped RS geometry. Yet another model~\cite{Csaki:2020zqz} used a similar idea of crunching away patches with large CC, by mixing the Higgs to the dilaton of a strongly coupled CFT (dual to radion in the RS framework), such that a meta stable minimum appears when the Higgs VEV is non-zero and below the TeV scale. As a result, only Hubble patches with unnaturally small values of the Higgs mass support inflation and post-inflationary expansion, while all other patches rapidly crunch. Interestingly, a consequence of this structure is again a very light radion. Although the verdict is out on these models, all these directions prompt thinking about the cosmological constant and model building more generally in a different way, tying it to EW sector and cosmology.


 

In addition to explaining the various hierarchies that appear in the SM, RS geometries can also be used to understand dynamics of hidden sectors. As an example, RS is used in ref.~\cite{Cesarotti:2020uod} to construct a phenomenological model of a generic strongly coupled hidden sector. The motivation for this kind of model stems from the necessity to discriminate a generic new physics signal event from the immense SM background. Since QCD primarily results in 2 or 3-jet events, a potentially lucrative and robust approach to identifying new physics is to search for signature of many isotropically distributed jets. However, generating template events that look quasi-isotropic can be poorly-motivated or intractable due to the number of free parameters in the toy model. The authors of ref.~\cite{Cesarotti:2020uod} therefore develop a 5d simplified model using RS framework that can generate spherical radiation patterns with a minimal number of parameters. 
\begin{figure}
    \centering
    \includegraphics[width=0.32\textwidth]{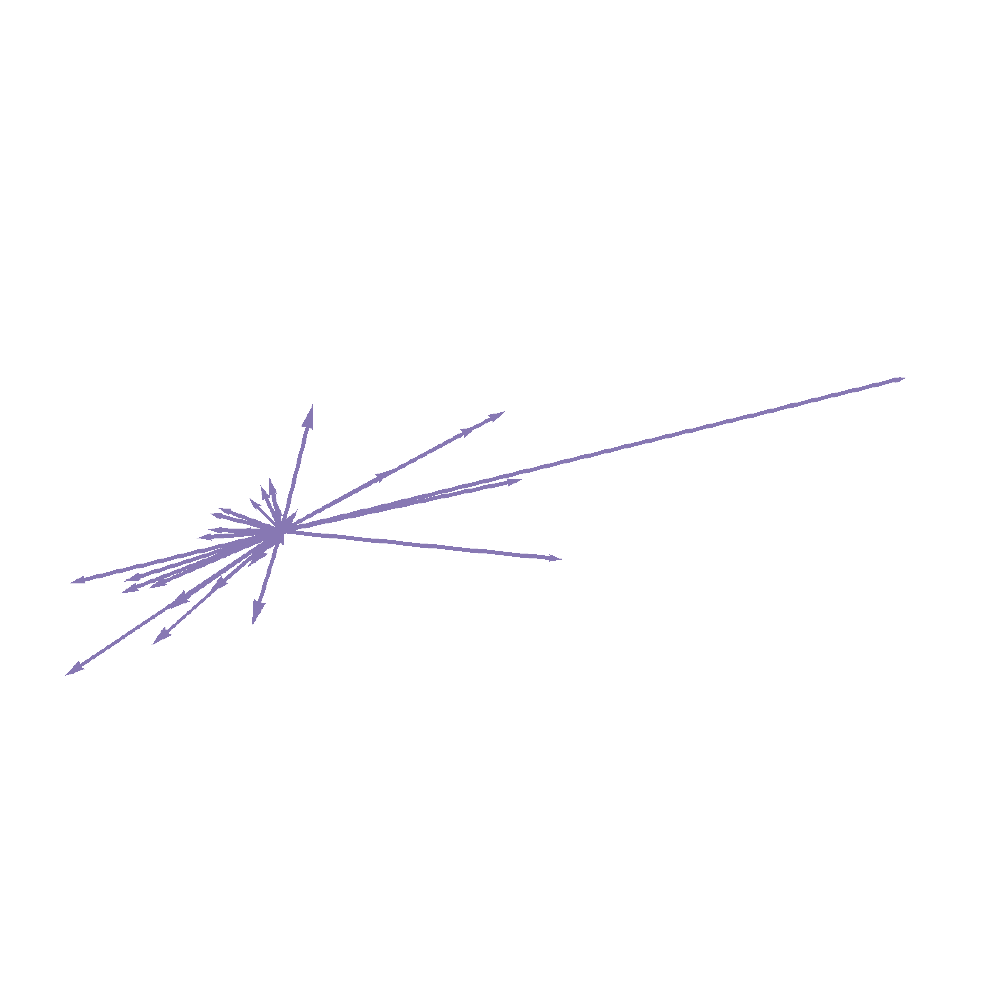}
    \includegraphics[width=0.32\textwidth]{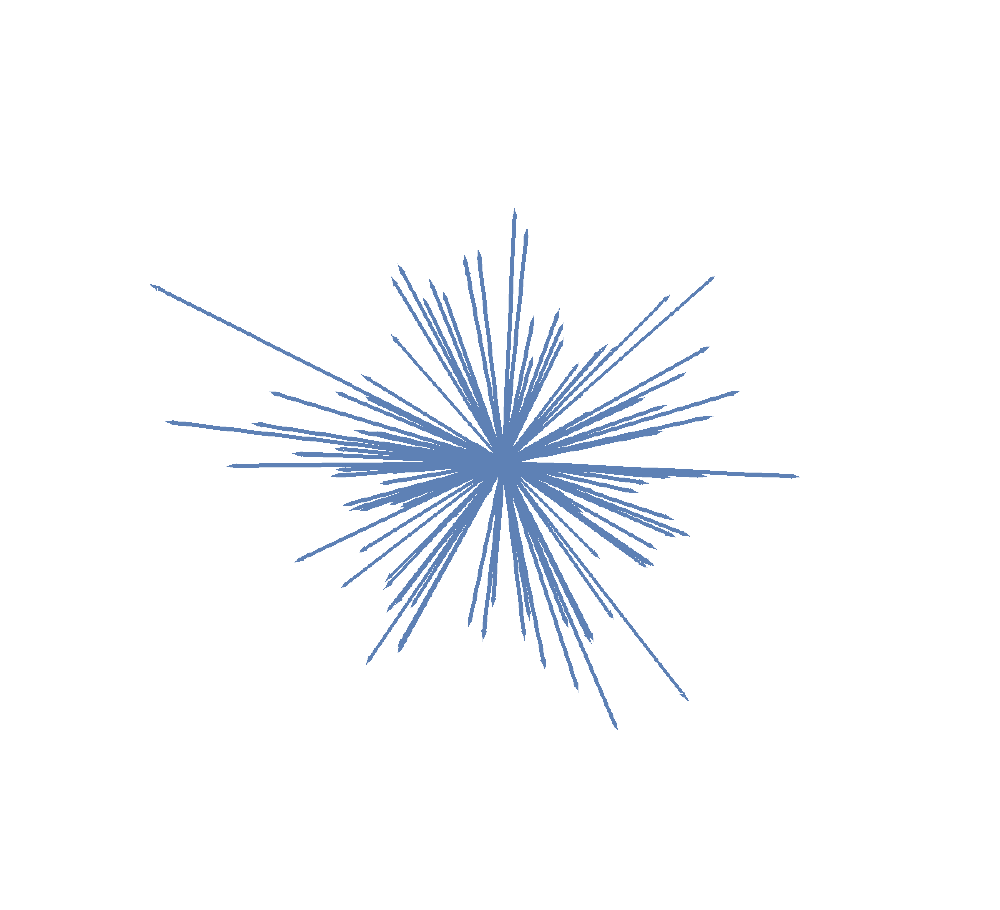}
    \includegraphics[width=0.32\textwidth]{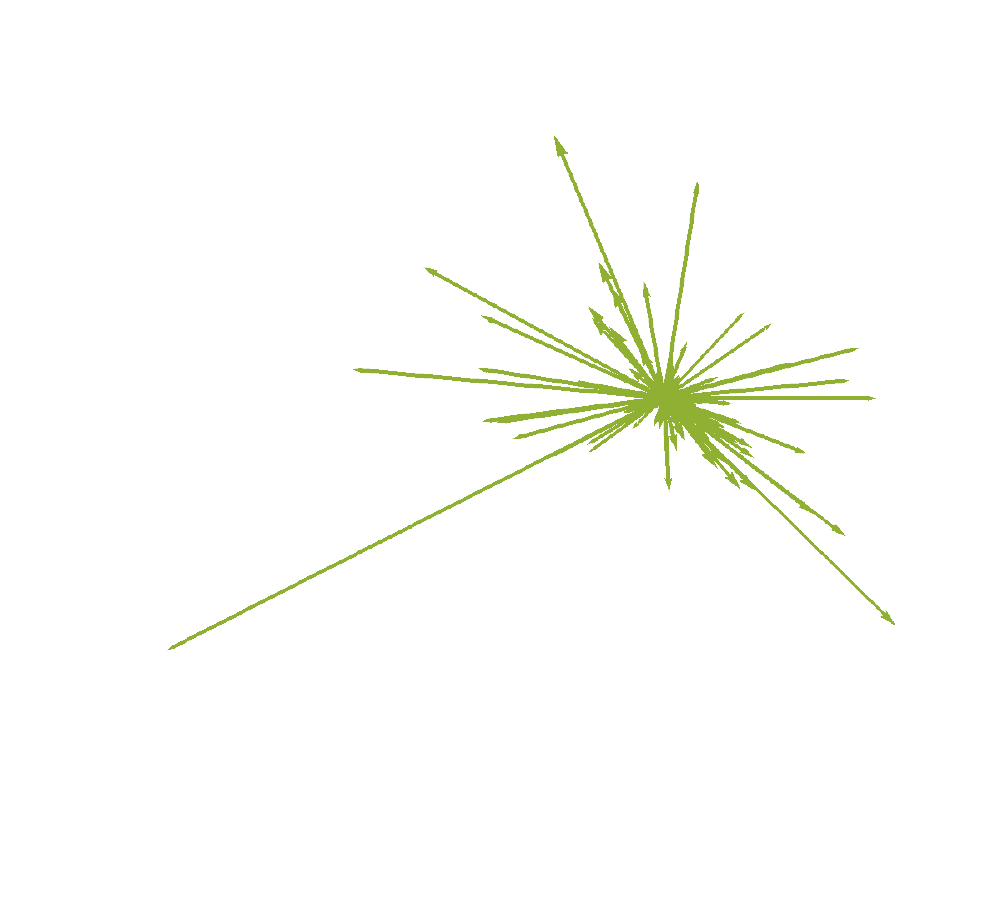}
    \caption{\small{Visualizations of a characteristic final state radiation patterns produced with the 5d simplified model framework. The length of the vectors corresponds to the total energy of the particle. By varying the input parameters, the simulation can produce collimated (left), uniform (center), and intermediate (right) event shapes. Figures adapted from ref.~\cite{Cesarotti:2020uod}.} }
    \label{fig:suepRS}
\end{figure}

This approach is motivated by the appearance of KK towers in extra dimensions which introduce many new particles to the model without requiring a large number of parameters. The curvature in RS is essential for cascade decays in the model to proceed in a quasi-isotropic pattern: a flat extra dimension would yield decays exactly at threshold, producing a high-multiplicity cloud of particles at rest; the curvature allows slightly off-threshold, therefore unboosted with randomly oriented non-zero momentum. The global event shape can be tuned via the 5d scalar mass and interpolates between a wide array of topologies, as shown in \fig{fig:suepRS}. Ultimately this 5d simplified model approach produces a minimal toy model to study spherical radiation patterns. Furthermore, this work has also been applied to develop and quantify a novel event shape observable, event isotropy~\cite{Cesarotti:2020ngq}, which has improved dynamic range in the quasi-isotropic regime. It is important to note that this is one particular study of the dynamics generated within the RS framework, and that many other phenomenological models can stem from warped extra-dimensional geometry to study novel cascades in hidden sectors~\cite{Dienes:2018bbv, Brax:2019koq, Costantino:2020msc}, and motivate thinking about portals to hidden sectors~\cite{Contino:2020tix} in model-independent ways.

\section{Cosmology}

The cosmological history of the universe provides a way to probe very high energy dynamics, and has become increasingly relevant as we enter the era of gravitational wave (GW) detection and precision cosmology. GWs can probe phase transitions in the  early universe before the CMB epoch, and have the potential to transmit vital information about the degrees of freedom and the dynamics of the phase transitions. The RS model provides a rich playground for novel cosmological signals. This is an exciting arena with several open questions in phenomenology and  model building, and deep questions in quantum field theory such as the confinement-deconfinement transition. This has led to a very active ongoing exploration of cosmology in RS models with many open questions that we describe in this section.
The high temperature phase of the RS model is described by a 5-dimensional black hole which can have significant impact on the mechanisms to populate cosmological relics that rely on a global symmetry. The phase transition to the low-temperature phase is generically first-order and occurs around the TeV scale, producing GWs peaked at mHz frequency, which is exactly in the sensitivity range of upcoming  space-based GW measurements, such as those expected from  LISA~\cite{Randall:2006py}.


The holographic dual of the 5D geometry in the RS model is expected to be a nearly-conformal gauge theory confining at around the TeV scale. Thus we expect a confinement transition in RS at high temperatures, analogous to the QCD confinement crossover at $T \sim 100$ MeV. 
The seminal work of ref.~\cite{Witten:1998zw} laid the groundwork for studying confining phase transitions in holographic theories. This framework was applied more explicitly to the RS phase transition in ref.~\cite{Creminelli:2001th}.
As the universe gets reheated after the end of inflation, the thermodynamically preferred phase of the RS model is described by a background geometry in which the IR brane is replaced by a black brane--a horizon with translational symmetry along the transverse directions. The geometry is of a Schwarzschild blackhole in global-AdS,
referred to as the AdSS geometry.  In equilibrium, the location of the black brane is determined by the temperature of the universe. As the temperature drops, the black brane moves towards the AdS horizon (in the deep IR). Below a critical temperature $T_c$, the black brane moves past the putative location of the IR brane (whose location is decided by the Planck-Electroweak hierarchy), and the standard RS geometry becomes the thermodynamically preferred phase. The phase transition between these two phases is of first order, since the two geometries are both solutions of Einstein's equations and hence are local saddles of the path integral, thereby having a barrier separating them in the configuration space. By AdS/CFT duality, these two phases are described by the confined and deconfined phase of the dual CFT. 

The details of the phase transition are hard to calculate in general, but can be estimated in models where the radion remains light~\cite{Creminelli:2001th}. The transition occurs through the nucleation of a true vacuum bubble in the AdSS phase, where the RS geometry inside connects with the AdSS geometry surrounding it with a gravitational instanton. The tunneling exponent can be then computed under the assumption that the action of the bubble can be well-approximated by the radion contribution.
Even when the RS geometry becomes the stable state of the theory, the rate of phase transition $\Gamma$ (probability per unit four-volume) of transition to this state is heavily suppressed, and schematically scales as
\begin{align}
    \Gamma \sim \text{exp}\left(-\frac{M_5^3/k^3}{\delta}\right) \sim e^{-N_c^2/\delta}\:,
    \label{eq:RateForTunneling}
\end{align}
where $M_5$ is the 5D Planck mass, $k$ is the AdS curvature and $\delta$ is a dimensionless parameter related to the stabilization mechanism of the RS geometry. By AdS/CFT duality, it can be written as the second term in eq.~\eqref{eq:RateForTunneling}, where $N_c$ is the number of colors of the dual CFT, and $\delta$ is now a parameter related to an explicit breaking of the CFT. For perturbative control of the 5D description, we need $(M_5/k)^3 \sim N_c^2 \gg 1$, while the requirement of stabilizing the geometry with a large hierarchy requires $\delta \lesssim 1$, both of which reduce the rate of transition exponentially. The requirement for a phase transition at $T\lesssim T_c$ translates to a very stringent bound on $M_5/k$:
\begin{align}
    (M_5/k)^3 \lesssim \delta \log (M_\text{Pl}/T_c)\:,
\end{align}
where $M_\text{Pl}$ is the 4D Planck scale. To explain the electroweak hierarchy, $T_c \sim \mathcal{O}(\text{TeV})$ which translates to $M_5/k \lesssim 10$, even for $\delta \sim \mathcal{O}(1)$ - the low energy gravitational description being barely under theoretical control. The way out is to require the universe to undergo some amount of supercooling so that the phase transition completes at $T_n \ll T_c$, and the rate for transition is enhanced by some function of the ratio $T_c/T_n$. However, this barely helps with making $M_5/k$ parametrically large, and further, $T_n$ can not be arbitrarily small if this description is to fit the standard BBN picture. 

Assuming these issues are addressed in more complete models (such as those referenced below), the features of being a first order phase transition and requiring supercooling can have dramatic observational consequences. The first order nature of the phase transition can lead to gravitational wave (GW) emission~\cite{Randall:2006py,vonHarling:2017yew,Baratella:2018pxi,Agashe:2020lfz,Baldes:2021aph}
which can be observed in present and upcoming GW experiments such as LISA~\cite{Caprini:2015zlo,Caprini:2019egz}, DECIGO~\cite{Seto:2001qf} and BBO~\cite{Corbin:2005ny}. This is an exciting opportunity on many frontiers -- as the theoretical and experimental advances in the GW detection program develop over coming decade, the opportunity to probe early universe dynamics in this BSM framework will become increasingly relevant. The supercooling feature affects the relic density of the relics of the cosmic past since there is a period for which the universe inflates, driven by the vacuum energy in the deconfined phase, and dilutes all the relic densities. This opens up parts of parameter space hitherto ruled out in the standard scenario.
This is a further opportunity for novel model building and exploration of theories and mechanisms that become relevant in this context.

The theoretical activities of the past decade have focused broadly on how robust this feature of a suppressed rate of phase transition and the associated supercooling is. There are two broad categories under which the recent approaches can be classified. 
\subsection*{IR modification}
The suppression from the small parameter $\delta$ in eq.~\eqref{eq:RateForTunneling} is tied to the shape of the radion potential. Close to the temperatures at which the phase transition becomes  feasible, a change in the radion potential  can modify the estimate for the rate of the phase transition. The idea of ref.~\cite{vonHarling:2017yew} was to use the corrections to the radion potential coming from the QCD condensate, which reduced the barrier depth, and hence enhanced the rate for the phase transition. The authors in~\cite{Baratella:2018pxi} further required the 4D CFT degrees of freedom to flow to another fixed point after QCD confinement, thereby gaining more theoretical control, and gave a geometrical picture for this situation where the mass of the GW scalar is tachyonic~\cite{Pomarol:2019aae}. The more recent work in~\cite{Agashe:2019lhy, Agashe:2020lfz} used similar ingredients where they modelled the IR flow to another fixed point by a special form of the potential for the Goldberger-Wise (GW) field, without the need for QCD corrections to the low energy radion EFT.

%
%

\subsection*{High Temperature modification} 
An alternative approach to the phase transition is to modify the description of the theory at high temperatures. The authors of ref.~\cite{Agrawal:2021alq} considered additional scalars in the low energy EFT apart from the radion, and calculated finite temperature corrections to the combined potential. The resulting potential has a meta-stable minimum at high temperatures, whose location moves with temperature, eventually merging to the temperature independent minimum in a standard radion potential. 
Additionally requiring that the state of the universe after preheating is not AdSS but rather the modified RS, the need of a phase transition and the issue of a small rate are avoided altogether, at the cost of the initial condition. The model is stable against a reverse phase transition to the AdSS phase, simply because of the suppressed rate, turning this feature into an advantage. 

The observational consequences of this modification are striking. Early universe cosmological phenomena such as WIMP freeze-out, axion abundance, baryogenesis, phase transitions, and gravitational wave signatures are qualitatively modified. In particular, the temperature dependent KK scale, to be contrasted with the IR modifications, has novel consequences, which were pointed out in ref.~\cite{Agrawal:2021alq}.

\subsection*{Future directions}
 Several interesting questions remain if we are to better understand the range of cosmological possibilities. The role of the radion and the related low energy EFT, on which the calculations for the rate are based, has not been fully explored. In particular, the presence of other light fields in the EFT can change the picture significantly. Other warped geometries such as Klebanov-Tseytlin (KT) and Klebanov-Strassler (KS) geometries can also change the description significantly, both by a different low energy EFT description, as well as a richer high temperature phase (corresponding to the AdSS geometry in simplified description) and symmetry breaking patterns~\cite{Buchel:2021yay}. Perhaps more drastically, a better understanding of low energy dynamics derived from warped compactifications might be required.

The observational consequences of alternative scenarios is another fruitful direction. We already saw that the cosmological history and the relic abundances are very different in the UV and IR modifications that have been considered in the literature. A more detailed modeling of the GW signal can tell us more about these scenario. Further exploration of these ideas can significantly broaden our understanding of possible cosmological histories consistent with present observations, and may give us a peek into the UV. 

\section{Warped Geometry  in String Theory} 

\begin{figure}[ht]
\begin{subfigure}{.5\textwidth}
  \centering
  \includegraphics[scale=0.37]{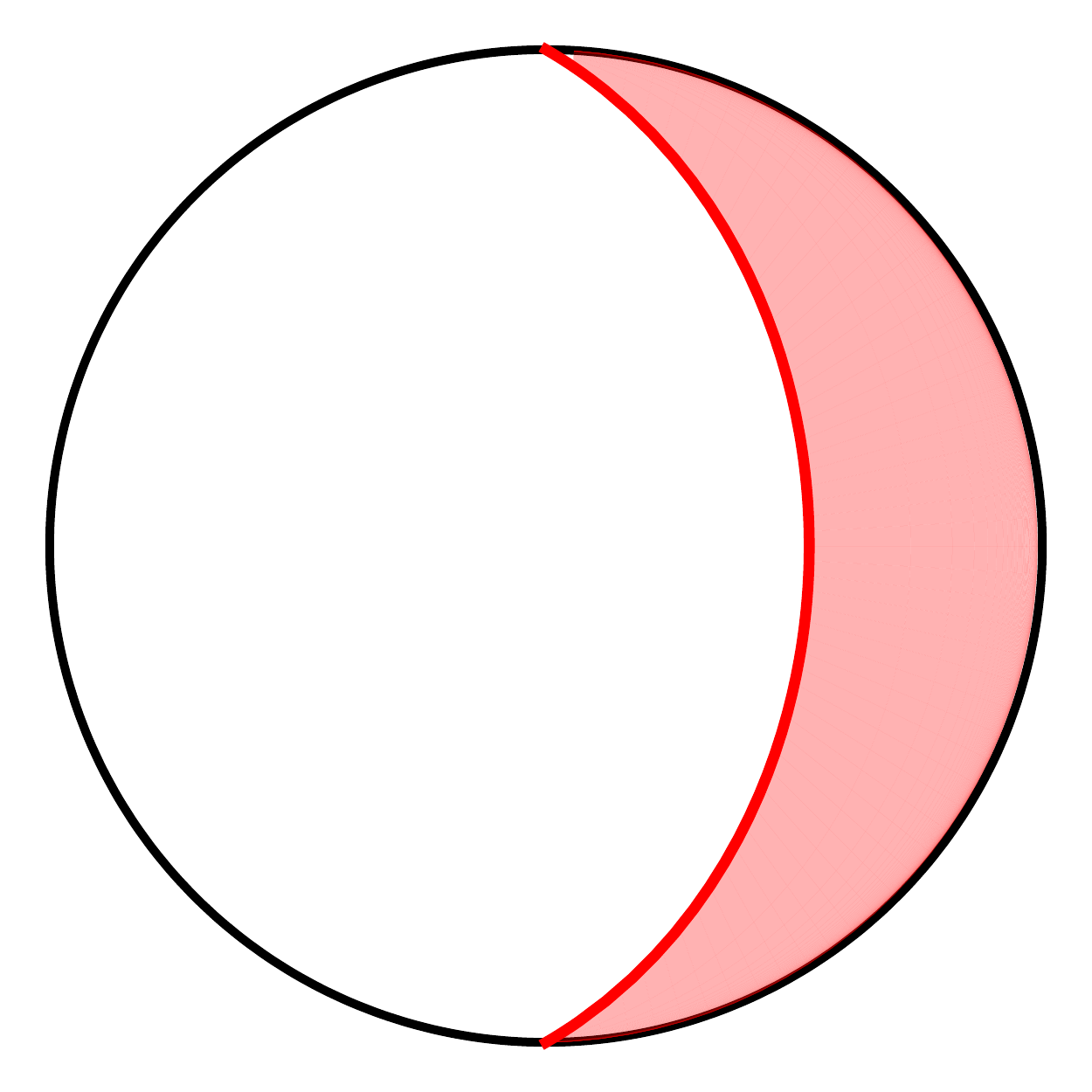}  
  \caption{\small{Vertical Cross Section}}
  \label{adstop}
\end{subfigure}
\begin{subfigure}{.5\textwidth}
  \centering
  \includegraphics[scale=0.3]{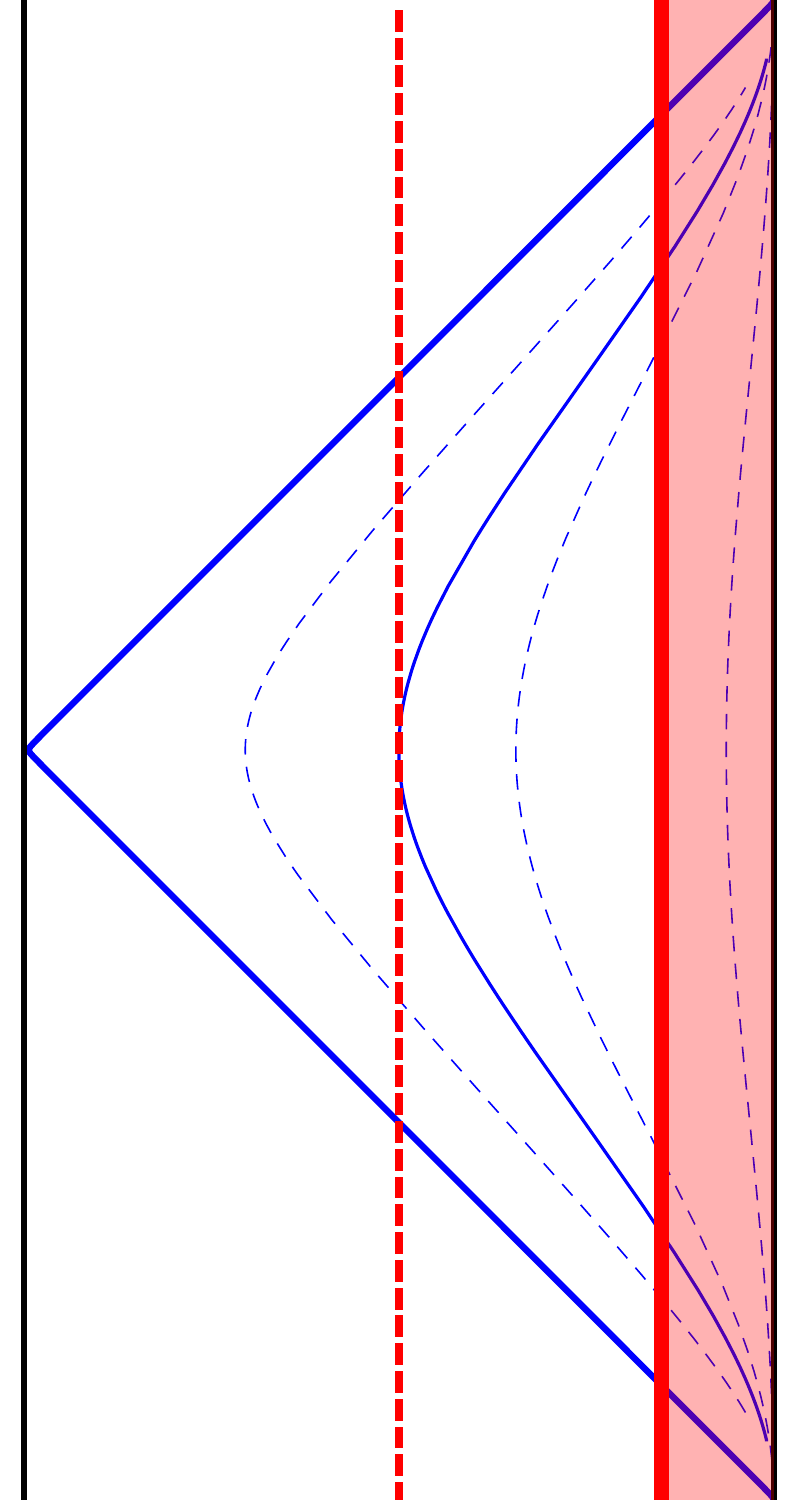}  
  \caption{\small{Horizontal Cross Section}}
\end{subfigure}
\caption{\small{Horizontal and vertical cross section of global Anti-de Sitter space with embedded branes realizing a BCFT. The red brane itself also has an Anti-de Sitter worldvolume the (red) shaded areas are excised by the brane.}}
\label{fig:adsbcft}
\end{figure}

Warped geometries with branes have found important applications in addressing fundamental questions in quantum gravity and string theory. Crucial for this connection is the fact that these braneworlds naturally result in geometries that are locally AdS space and so can be interpreted with the AdS/CFT correspondence. The latter asserts that gravity on AdS is equivalent to a conformal field theory living on its boundary, providing a setting in which quantum gravity can be reliably studied. This connection to AdS/CFT allows an interpretation of warped compactifications with branes in a dual field theory language with interesting consequences (non-perturbative RG, compositeness). One application that was recognized early on is that this connection provides a generalization of AdS/CFT to spaces with boundaries \cite{Karch:2000ct,Karch:2000gx}.  As a direct offshoot of this investigation, the first example of a consistent theory of interacting massive gravity was uncovered which paved the way for many of the advances that followed in this arena. This holographic duality between conformal field theories with boundaries and warped geometries with branes has since then been significantly deepened in \cite{Takayanagi:2011zk,Fujita:2011fp}, where it was advanced as a general duality between AdS and boundary conformal field theories, AdS/BCFT for short. The basic geometry of a braneworld dual to a BCFT is summarized in figure \ref{fig:adsbcft}. 

%
%

Given the interesting aspects of warped geometries that have been uncovered in terms of effective theories of general relativity coupled to brane models,  efforts soon followed to understand these ``bottom-up" models from the ``top-down". The main purpose of these studies was to embed warped compactifications in the framework of string theory. Ref. \cite{Verlinde:1999fy} laid out the earliest ideas of how to realize warped compactifications with Minkowski slicing in string theory, where it was argued that stacks of branes that are required to cancel tadpoles arising from  flux sources in string theory can pull out an AdS-like throat from an otherwise compact manifold, giving essentially a realization of a UV brane.  A subsequent important development was recognizing other types of warped geometries that have  natural embedding in string theory, such as that proposed by Klebanov and Strassler \cite{Klebanov:2000hb}; these warped throats nicely allowed internal geometries to truncate, giving a top-down realization of an IR brane. Ref \cite{Giddings:2001yu} formalized the framework of warped compactifications with fluxes, realizing naturally large hierarchies,  which served as the starting point for many interesting string constructions over the years. From the string theory point of view, a very different route led to the realization of top-down AdS/BCFT models within string theory. Here early suggestions for how to realize a BCFT in string theory from a system of D3 branes ending on 5-branes was made in \cite{Karch:2001cw}, with a concrete realization of these ideas being worked out in a series of papers \cite{DHoker:2007hhe,DHoker:2007zhm} several years later. The techniques of this latter work have since then laid the foundation for constructing string theory embeddings of supersymmetric boundary conformal field theories in a variety of dimensions and with varying amount of supersymmetry, as laid out in dozens of papers.

Significant progress has also been made 
in the other direction: using branes within string theory to aid in the construction of novel string theory models--most notably in helping generate a tunable vacuum energy that can be used for de Sitter models.   The observation of a positive cosmological vacuum energy is a big challenge in string theory in that the only controlled models that resemble our world rely on spacetime supersymmetry, which can never have a positive energy vacuum. Breaking supersymmetry in a controlled fashion when the associated energy is far less than that of electroweak physics introduces a big challenge. One way to address this is to generate a small parameter through warping—one that competes with a supersymmetry breaking energy that is suppressed by mass scales. Most prominent among these attempts is the development of controlled models of de Sitter space, such as the one famously pioneered by Kachru, Kallosh, Linde, and Trivedi \cite{Kachru:2003sx}. Many details of these construction have since been fleshed out with 
progress towards an understanding of the 10d solutions, appearing 
in \cite{Kachru:2019dvo,Demirtas:2021nlu}. This theory has introduced a good deal of interest and controversy and has in part catalyzed the swampland studies which are quite prevalent now  and will continue to dominate some studies of string vacua. While the full solutions of this type appear complicated at first sight, many of their essential properties can be understood and systematically worked out in a low energy effective description in terms of 5d warped geometries with branes \cite{Randall:2019ent}.

A different route to de Sitter space built on the idea of warped compactifications with branes is provided by the dS/dS correspondence \cite{Alishahiha:2004md}. Very similar to the AdS realization of BCFTs, which are based on AdS spaces sliced in a warped compactification via lower dimensional AdS space, the dS/dS correspondence is built on a slicing of dS space by dS slices. It was shown in \cite{Karch:2003em} that, when viewed as a warped compactification, de Sitter space is ``autolocalizing" -- it has effectively a built in UV brane even in the absence of any brane sources. Based on this observation, the dS/dS correspondence provides a dual description of de Sitter gravity in terms of lower de Sitter gravity by interpreting it as a brane world with localized gravity. Based on this idea, string theory construction of de Sitter spaces based on intersecting brane models can be given \cite{Dong:2010pm}. More recently, this duality has been put on a firm footing using the tool of the so called TTbar deformation \cite{Gorbenko:2018oov}. These latter ideas have since then been applied to work towards a holographic understanding of de Sitter space also directly in the static patch \cite{Coleman:2021nor}.


\section{Applications to the black hole information paradox}
The particular case of warped AdS branes has also  found interesting applications in addressing questions about black hole evaporation in a fully realized setting. 
Recent developments  using quantum information tools to understand the recovery of information from an evaporating black hole have led to remarkable progress, most prominent among them the calculation of the so called Page curve, describing the initial loss of information and its subsequent recovery.
The basic story of the Page curve calculation was first developed in the context of low dimensional gravity. Using the concept of a quantum entangling surface, a Page curve for an evaporating black hole consistent with unitarity was obtained using semi-classic methods. Much of this progress is summarized in a separate white paper dedicated to this topic \cite{Bousso:2022ntt}. While claims have been put forward that this mechanism may be more widely applicable, most concrete calculations have been performed in the context of gravity coupled to an external non-gravitating bath. How important this coupling to the bath is is a topic of active debate. In fact, recent work suggest that the non-trivial Page curve disappears if the bath is removed and is replaced by an entirely trivial flat Page curve \cite{Laddha:2020kvp,Raju:2021lwh,Geng:2021hlu}.
 
Braneworlds, which in this context are often referred to as end-of-the-world branes, naturally provide a setting in which AdS gravity can be coupled to an external bath, allowing the study of black hole evaporation in the well-controlled setting of AdS/CFT.  
In fact, all examples of Page curve calculations in more than 1+1 dimensions (starting with the seminal paper \cite{Almheiri:2019psy}), as well as some of the lower dimensional constructions \cite{Almheiri:2019hni}, are based on the realization of gravity localized on a brane in Anti-de Sitter space. This includes the first embedding into full fledged string theory of a unitary Page curve in more than 1+1 dimensions \cite{Uhlemann:2021nhu}. As already described in the first papers on warped AdS branes \cite{Karch:2000ct,Karch:2000gx}, one has three dual descriptions of the system in this case: the gravitational description in terms of classical Einstein equations with a brane source, which can be solved classically, the BCFT dual, which in principle gives a UV complete quantum mechanical description of the system, as well as the one of interest for black hole evaporation: the ``intermediate" picture of gravity on the brane coupled to the bath represented by the BCFT ambient space. This double-holographic dictionary allows many new perspectives on the evaporation problem. For example, the emergence of so called islands in the resolution of the black hole information paradox can be understood from a purely BCFT point of view as a change of dominance of two different exchange channels in the OPE of the BCFT \cite{Rozali:2019day}.

One very interesting feature of these particular realizations of evaporating black holes on branes is that the setups automatically yield a theory of massive gravity, as recently emphasized in~\cite{Geng:2020qvw}. The mass of the graviton was first found from the point of view of the classical higher dimensional gravity from solving the small fluctuation problem in brane sourced Einstein equations in \cite{Karch:2000ct,Karch:2000gx} and was understood in terms of a loop induced mass for the graviton from the boundary conditions connecting to the bath by Porrati in \cite{Porrati:2001db}. From the BCFT point of view, the graviton mass is given by the anomalous dimension of the boundary stress tensor, which is not conserved due to the interactions with the BCFT ambient space \cite{Aharony:2003qf}. In fact, this field theory explanation makes it clear that a graviton mass in an unavoidable consequence of coupling gravity to any bath \cite{Aharony:2006hz,Kiritsis:2006hy}. Whether this mass is a crucial part of the Page curve story, as argued in \cite{Geng:2020qvw,Geng:2020fxl,Geng:2021hlu}, or just a technicality that has no bearing on the information recovery from the black hole, is a topic under active investigation. No matter the ultimate answer, warped brane-worlds will be one of the crucial tools we have in order to address this question.

While the main role of the end of the world branes in these recent black hole studies has been to realize AdS gravity coupled to a bath, they have also played a separate role in at least one of the most studied toy models of black hole evaporation: the west coast model \cite{Penington:2019kki}. In this work, end of the world branes were not merely the canvas on which the gravitational universe lives, they were models for the matter that lives inside the black hole, realizing some of the micro-states of the black hole. 

Besides giving simple existence proofs of unitarily evaporating black holes in higher dimensions, warped brane worlds have also been used to address other aspects of this problem. In many cases, an interesting phase structure exists on when and how entanglement islands form in order to unitarize the theory. These phase structures can easily be mapped out by performing calculations using the classical bulk gravity \cite{Chen:2020hmv,Chen:2020uac,Geng:2021mic}.

In addition to using brane worlds as a tool to study the black hole information paradox, recent work has also used it to study parallel information paradoxes in cosmology~\cite{Geng:2021wcq}, and also drawn attention to the fact that several fundamental aspects of the brane worlds themselves, in particular their holographic dictionary, are still poorly understood. Only very recently has an explicit dictionary defining the intermediate picture in terms of bulk quantities been proposed \cite{Neuenfeld:2021wbl}. Furthermore, a very intriguing apparent non-locality in the intermediate picture has been pointed out in \cite{Omiya:2021olc}. Clearly these questions deserve further investigation.

In an orthogonal but somewhat related recent development, RS2 modeling has been invoked as a key feature in the construction of smooth traversable wormhole solutions~\cite{Maldacena:2020sxe}. The subject of asymptotic symmetries has also enjoyed renewed interest in the last several years. In this context, novel infinite-dimensional asymptotic symmetries of gauge and gravitational theories in 4D asymptotically AdS spacetime were discovered by exploiting an RS2-like construction~\cite{Mishra:2017zan}. In a weak gauging limit, the AdS correlators can be related to those of a 2D CFT, defined at the intersection of the asymptotic boundary and a probe brane. This makes the underlying Virasoro and Kac-Moody Ward identities manifest. The connection of this construction to questions about information paradox, and the relevance to the existence of a Page curve, is an interesting direction to pursue.

\section{Summary}
In this white paper, we have presented a brief sampling of some of the important implications of warped geometry and its underlying resurgence and persistence in recent years. Both theoretically and phenomenologically this framework yields tractable testable examples that can be rigorously studied in detail. Warped geometries has had and will continue to have an enormous impact on theoretical particle physics.

\section*{Acknowledgements}

We thank Hao Geng and Shamit Kachru for useful discussions and comments on the draft.

\noindent\rule{\textwidth}{1pt}

\bibliographystyle{utphys}
\bibliography{references}

\providecommand{\href}[2]{#2}\begingroup\raggedright\begin{thebibliography}{10}

\bibitem{Almheiri:2020cfm}
A.~Almheiri, T.~Hartman, J.~Maldacena, E.~Shaghoulian, and A.~Tajdini, ``{The
  entropy of Hawking radiation},''
  \href{http://dx.doi.org/10.1103/RevModPhys.93.035002}{{\em Rev. Mod. Phys.}
  {\bfseries 93} no.~3, (2021) 035002},
  \href{http://arxiv.org/abs/2006.06872}{{\ttfamily arXiv:2006.06872
  [hep-th]}}.

\bibitem{Agashe:2004cp}
K.~Agashe, G.~Perez, and A.~Soni, ``{Flavor structure of warped extra dimension
  models},'' \href{http://dx.doi.org/10.1103/PhysRevD.71.016002}{{\em Phys.
  Rev. D} {\bfseries 71} (2005) 016002},
  \href{http://arxiv.org/abs/hep-ph/0408134}{{\ttfamily arXiv:hep-ph/0408134}}.

\bibitem{Cacciapaglia:2007fw}
G.~Cacciapaglia, C.~Csaki, J.~Galloway, G.~Marandella, J.~Terning, and
  A.~Weiler, ``{A GIM Mechanism from Extra Dimensions},''
  \href{http://dx.doi.org/10.1088/1126-6708/2008/04/006}{{\em JHEP} {\bfseries
  04} (2008) 006}, \href{http://arxiv.org/abs/0709.1714}{{\ttfamily
  arXiv:0709.1714 [hep-ph]}}.

\bibitem{Fitzpatrick:2007sa}
A.~L. Fitzpatrick, G.~Perez, and L.~Randall, ``{Flavor anarchy in a
  Randall-Sundrum model with 5D minimal flavor violation and a low Kaluza-Klein
  scale},'' \href{http://dx.doi.org/10.1103/PhysRevLett.100.171604}{{\em Phys.
  Rev. Lett.} {\bfseries 100} (2008) 171604},
  \href{http://arxiv.org/abs/0710.1869}{{\ttfamily arXiv:0710.1869 [hep-ph]}}.

\bibitem{Perez:2008ee}
G.~Perez and L.~Randall, ``{Natural Neutrino Masses and Mixings from Warped
  Geometry},'' \href{http://dx.doi.org/10.1088/1126-6708/2009/01/077}{{\em
  JHEP} {\bfseries 01} (2009) 077},
  \href{http://arxiv.org/abs/0805.4652}{{\ttfamily arXiv:0805.4652 [hep-ph]}}.

\bibitem{Agashe:2015izu}
K.~Agashe, S.~Hong, and L.~Vecchi, ``{Warped seesaw mechanism is physically
  inverted},'' \href{http://dx.doi.org/10.1103/PhysRevD.94.013001}{{\em Phys.
  Rev. D} {\bfseries 94} no.~1, (2016) 013001},
  \href{http://arxiv.org/abs/1512.06742}{{\ttfamily arXiv:1512.06742
  [hep-ph]}}.

\bibitem{flavorSnowmass2021}
W.~Altmannshofer and J.~Zupan, ``{Snowmass 2021 White Paper: Flavor model
  building},'' {\em Contribution to Snowmass 2021} (2022) .

\bibitem{Agashe:2016rle}
K.~Agashe, P.~Du, S.~Hong, and R.~Sundrum, ``{Flavor Universal Resonances and
  Warped Gravity},'' \href{http://dx.doi.org/10.1007/JHEP01(2017)016}{{\em
  JHEP} {\bfseries 01} (2017) 016},
  \href{http://arxiv.org/abs/1608.00526}{{\ttfamily arXiv:1608.00526
  [hep-ph]}}.

\bibitem{Agashe:2016kfr}
K.~S. Agashe, J.~Collins, P.~Du, S.~Hong, D.~Kim, and R.~K. Mishra, ``{LHC
  Signals from Cascade Decays of Warped Vector Resonances},''
  \href{http://dx.doi.org/10.1007/JHEP05(2017)078}{{\em JHEP} {\bfseries 05}
  (2017) 078}, \href{http://arxiv.org/abs/1612.00047}{{\ttfamily
  arXiv:1612.00047 [hep-ph]}}.

\bibitem{Agashe:2017wss}
K.~Agashe, J.~H. Collins, P.~Du, S.~Hong, D.~Kim, and R.~K. Mishra,
  ``{Dedicated Strategies for Triboson Signals from Cascade Decays of Vector
  Resonances},'' \href{http://dx.doi.org/10.1103/PhysRevD.99.075016}{{\em Phys.
  Rev. D} {\bfseries 99} no.~7, (2019) 075016},
  \href{http://arxiv.org/abs/1711.09920}{{\ttfamily arXiv:1711.09920
  [hep-ph]}}.

\bibitem{Agashe:2018leo}
K.~Agashe, J.~H. Collins, P.~Du, S.~Hong, D.~Kim, and R.~K. Mishra,
  ``{Detecting a Boosted Diboson Resonance},''
  \href{http://dx.doi.org/10.1007/JHEP11(2018)027}{{\em JHEP} {\bfseries 11}
  (2018) 027}, \href{http://arxiv.org/abs/1809.07334}{{\ttfamily
  arXiv:1809.07334 [hep-ph]}}.

\bibitem{CMS:2021nrj}
``{Search for resonances decaying to triple W-boson final states in
  proton-proton collisions at $\sqrt{s}=13~\mathrm{TeV}$}, cms-pas-b2g-20-001,
  2021.,''.

\bibitem{CMS:2021qyp}
``{Search for $\mathrm{W^\prime}$ decaying to a vector-like quark and a top or
  bottom quark in the all-jets final state, CMS-PAS-B2G-20-002, 2021},''.

\bibitem{CMS:2021qev}
``{Search for high mass trijet resonances using final states with boosted dijet
  resonances in proton-proton collisions at $\sqrt{s}=13~\mathrm{TeV}$,
  CMS-PAS-EXO-20-007, 2021.},''.

\bibitem{Csaki:2018kxb}
C.~Cs\'aki, G.~Lee, S.~J. Lee, S.~Lombardo, and O.~Telem, ``{Continuum
  Naturalness},'' \href{http://dx.doi.org/10.1007/JHEP03(2019)142}{{\em JHEP}
  {\bfseries 03} (2019) 142}, \href{http://arxiv.org/abs/1811.06019}{{\ttfamily
  arXiv:1811.06019 [hep-ph]}}.

\bibitem{Arvanitaki:2016xds}
A.~Arvanitaki, S.~Dimopoulos, V.~Gorbenko, J.~Huang, and K.~Van~Tilburg, ``{A
  small weak scale from a small cosmological constant},''
  \href{http://dx.doi.org/10.1007/JHEP05(2017)071}{{\em JHEP} {\bfseries 05}
  (2017) 071}, \href{http://arxiv.org/abs/1609.06320}{{\ttfamily
  arXiv:1609.06320 [hep-ph]}}.

\bibitem{Bloch:2019bvc}
I.~M. Bloch, C.~Cs\'aki, M.~Geller, and T.~Volansky, ``{Crunching away the
  cosmological constant problem: dynamical selection of a small $\Lambda$},''
  \href{http://dx.doi.org/10.1007/JHEP12(2020)191}{{\em JHEP} {\bfseries 12}
  (2020) 191}, \href{http://arxiv.org/abs/1912.08840}{{\ttfamily
  arXiv:1912.08840 [hep-ph]}}.

\bibitem{Csaki:2020zqz}
C.~Cs\'aki, R.~T. D'Agnolo, M.~Geller, and A.~Ismail, ``{Crunching Dilaton,
  Hidden Naturalness},''
  \href{http://dx.doi.org/10.1103/PhysRevLett.126.091801}{{\em Phys. Rev.
  Lett.} {\bfseries 126} (2021) 091801},
  \href{http://arxiv.org/abs/2007.14396}{{\ttfamily arXiv:2007.14396
  [hep-ph]}}.

\bibitem{Cesarotti:2020uod}
C.~Cesarotti, M.~Reece, and M.~J. Strassler, ``{Spheres To Jets: Tuning Event
  Shapes with 5d Simplified Models},''
  \href{http://dx.doi.org/10.1007/JHEP05(2021)096}{{\em JHEP} {\bfseries 05}
  (2021) 096}, \href{http://arxiv.org/abs/2009.08981}{{\ttfamily
  arXiv:2009.08981 [hep-ph]}}.

\bibitem{Cesarotti:2020ngq}
C.~Cesarotti, M.~Reece, and M.~J. Strassler, ``{The efficacy of event isotropy
  as an event shape observable},''
  \href{http://dx.doi.org/10.1007/JHEP07(2021)215}{{\em JHEP} {\bfseries 07}
  (2021) 215}, \href{http://arxiv.org/abs/2011.06599}{{\ttfamily
  arXiv:2011.06599 [hep-ph]}}.

\bibitem{Dienes:2018bbv}
K.~R. Dienes, J.~Kost, and B.~Thomas, ``{Dynamics of Kaluza-Klein Towers in the
  Early Universe},'' \href{http://dx.doi.org/10.22323/1.340.0140}{{\em PoS}
  {\bfseries ICHEP2018} (2019) 140}.

\bibitem{Brax:2019koq}
P.~Brax, S.~Fichet, and P.~Tanedo, ``{The Warped Dark Sector},''
  \href{http://dx.doi.org/10.1016/j.physletb.2019.135012}{{\em Phys. Lett. B}
  {\bfseries 798} (2019) 135012},
  \href{http://arxiv.org/abs/1906.02199}{{\ttfamily arXiv:1906.02199
  [hep-ph]}}.

\bibitem{Costantino:2020msc}
A.~Costantino, S.~Fichet, and P.~Tanedo, ``{Effective Field Theory in AdS:
  Continuum Regime, Soft Bombs, and IR Emergence},''
  \href{http://dx.doi.org/10.1103/PhysRevD.102.115038}{{\em Phys. Rev. D}
  {\bfseries 102} no.~11, (2020) 115038},
  \href{http://arxiv.org/abs/2002.12335}{{\ttfamily arXiv:2002.12335
  [hep-th]}}.

\bibitem{Contino:2020tix}
R.~Contino, K.~Max, and R.~K. Mishra, ``{Searching for elusive dark sectors
  with terrestrial and celestial observations},''
  \href{http://dx.doi.org/10.1007/JHEP06(2021)127}{{\em JHEP} {\bfseries 06}
  (2021) 127}, \href{http://arxiv.org/abs/2012.08537}{{\ttfamily
  arXiv:2012.08537 [hep-ph]}}.

\bibitem{Randall:2006py}
L.~Randall and G.~Servant, ``{Gravitational waves from warped spacetime},''
  \href{http://dx.doi.org/10.1088/1126-6708/2007/05/054}{{\em JHEP} {\bfseries
  05} (2007) 054}, \href{http://arxiv.org/abs/hep-ph/0607158}{{\ttfamily
  arXiv:hep-ph/0607158}}.

\bibitem{Witten:1998zw}
E.~Witten, ``{Anti-de Sitter space, thermal phase transition, and confinement
  in gauge theories},''
  \href{http://dx.doi.org/10.4310/ATMP.1998.v2.n3.a3}{{\em Adv. Theor. Math.
  Phys.} {\bfseries 2} (1998) 505--532},
  \href{http://arxiv.org/abs/hep-th/9803131}{{\ttfamily arXiv:hep-th/9803131}}.

\bibitem{Creminelli:2001th}
P.~Creminelli, A.~Nicolis, and R.~Rattazzi, ``{Holography and the electroweak
  phase transition},''
  \href{http://dx.doi.org/10.1088/1126-6708/2002/03/051}{{\em JHEP} {\bfseries
  03} (2002) 051}, \href{http://arxiv.org/abs/hep-th/0107141}{{\ttfamily
  arXiv:hep-th/0107141}}.

\bibitem{vonHarling:2017yew}
B.~von Harling and G.~Servant, ``{QCD-induced Electroweak Phase Transition},''
  \href{http://dx.doi.org/10.1007/JHEP01(2018)159}{{\em JHEP} {\bfseries 01}
  (2018) 159}, \href{http://arxiv.org/abs/1711.11554}{{\ttfamily
  arXiv:1711.11554 [hep-ph]}}.

\bibitem{Baratella:2018pxi}
P.~Baratella, A.~Pomarol, and F.~Rompineve, ``{The Supercooled Universe},''
  \href{http://dx.doi.org/10.1007/JHEP03(2019)100}{{\em JHEP} {\bfseries 03}
  (2019) 100}, \href{http://arxiv.org/abs/1812.06996}{{\ttfamily
  arXiv:1812.06996 [hep-ph]}}.

\bibitem{Agashe:2020lfz}
K.~Agashe, P.~Du, M.~Ekhterachian, S.~Kumar, and R.~Sundrum, ``{Phase
  Transitions from the Fifth Dimension},''
  \href{http://dx.doi.org/10.1007/JHEP02(2021)051}{{\em JHEP} {\bfseries 02}
  (2021) 051}, \href{http://arxiv.org/abs/2010.04083}{{\ttfamily
  arXiv:2010.04083 [hep-th]}}.

\bibitem{Baldes:2021aph}
I.~Baldes, Y.~Gouttenoire, F.~Sala, and G.~Servant, ``{Supercool Composite Dark
  Matter beyond 100 TeV},'' \href{http://arxiv.org/abs/2110.13926}{{\ttfamily
  arXiv:2110.13926 [hep-ph]}}.

\bibitem{Caprini:2015zlo}
C.~Caprini {\em et~al.}, ``{Science with the space-based interferometer eLISA.
  II: Gravitational waves from cosmological phase transitions},''
  \href{http://dx.doi.org/10.1088/1475-7516/2016/04/001}{{\em JCAP} {\bfseries
  04} (2016) 001}, \href{http://arxiv.org/abs/1512.06239}{{\ttfamily
  arXiv:1512.06239 [astro-ph.CO]}}.

\bibitem{Caprini:2019egz}
C.~Caprini {\em et~al.}, ``{Detecting gravitational waves from cosmological
  phase transitions with LISA: an update},''
  \href{http://dx.doi.org/10.1088/1475-7516/2020/03/024}{{\em JCAP} {\bfseries
  03} (2020) 024}, \href{http://arxiv.org/abs/1910.13125}{{\ttfamily
  arXiv:1910.13125 [astro-ph.CO]}}.

\bibitem{Seto:2001qf}
N.~Seto, S.~Kawamura, and T.~Nakamura, ``{Possibility of direct measurement of
  the acceleration of the universe using 0.1-Hz band laser interferometer
  gravitational wave antenna in space},''
  \href{http://dx.doi.org/10.1103/PhysRevLett.87.221103}{{\em Phys. Rev. Lett.}
  {\bfseries 87} (2001) 221103},
  \href{http://arxiv.org/abs/astro-ph/0108011}{{\ttfamily
  arXiv:astro-ph/0108011}}.

\bibitem{Corbin:2005ny}
V.~Corbin and N.~J. Cornish, ``{Detecting the cosmic gravitational wave
  background with the big bang observer},''
  \href{http://dx.doi.org/10.1088/0264-9381/23/7/014}{{\em Class. Quant. Grav.}
  {\bfseries 23} (2006) 2435--2446},
  \href{http://arxiv.org/abs/gr-qc/0512039}{{\ttfamily arXiv:gr-qc/0512039}}.

\bibitem{Pomarol:2019aae}
A.~Pomarol, O.~Pujolas, and L.~Salas, ``{Holographic conformal transition and
  light scalars},'' \href{http://dx.doi.org/10.1007/JHEP10(2019)202}{{\em JHEP}
  {\bfseries 10} (2019) 202}, \href{http://arxiv.org/abs/1905.02653}{{\ttfamily
  arXiv:1905.02653 [hep-th]}}.

\bibitem{Agashe:2019lhy}
K.~Agashe, P.~Du, M.~Ekhterachian, S.~Kumar, and R.~Sundrum, ``{Cosmological
  Phase Transition of Spontaneous Confinement},''
  \href{http://dx.doi.org/10.1007/JHEP05(2020)086}{{\em JHEP} {\bfseries 05}
  (2020) 086}, \href{http://arxiv.org/abs/1910.06238}{{\ttfamily
  arXiv:1910.06238 [hep-ph]}}.

\bibitem{Agrawal:2021alq}
P.~Agrawal and M.~Nee, ``{Avoided deconfinement in Randall-Sundrum models},''
  \href{http://dx.doi.org/10.1007/JHEP10(2021)105}{{\em JHEP} {\bfseries 10}
  (2021) 105}, \href{http://arxiv.org/abs/2103.05646}{{\ttfamily
  arXiv:2103.05646 [hep-ph]}}.

\bibitem{Buchel:2021yay}
A.~Buchel, ``{A bestiary of black holes on the conifold with fluxes},''
  \href{http://dx.doi.org/10.1007/JHEP06(2021)102}{{\em JHEP} {\bfseries 06}
  (2021) 102}, \href{http://arxiv.org/abs/2103.15188}{{\ttfamily
  arXiv:2103.15188 [hep-th]}}.

\bibitem{Karch:2000ct}
A.~Karch and L.~Randall, ``{Locally localized gravity},''
  \href{http://dx.doi.org/10.1088/1126-6708/2001/05/008}{{\em JHEP} {\bfseries
  05} (2001) 008}, \href{http://arxiv.org/abs/hep-th/0011156}{{\ttfamily
  arXiv:hep-th/0011156}}.

\bibitem{Karch:2000gx}
A.~Karch and L.~Randall, ``{Open and closed string interpretation of SUSY CFT's
  on branes with boundaries},''
  \href{http://dx.doi.org/10.1088/1126-6708/2001/06/063}{{\em JHEP} {\bfseries
  06} (2001) 063}, \href{http://arxiv.org/abs/hep-th/0105132}{{\ttfamily
  arXiv:hep-th/0105132}}.

\bibitem{Takayanagi:2011zk}
T.~Takayanagi, ``{Holographic Dual of BCFT},''
  \href{http://dx.doi.org/10.1103/PhysRevLett.107.101602}{{\em Phys. Rev.
  Lett.} {\bfseries 107} (2011) 101602},
  \href{http://arxiv.org/abs/1105.5165}{{\ttfamily arXiv:1105.5165 [hep-th]}}.

\bibitem{Fujita:2011fp}
M.~Fujita, T.~Takayanagi, and E.~Tonni, ``{Aspects of AdS/BCFT},''
  \href{http://dx.doi.org/10.1007/JHEP11(2011)043}{{\em JHEP} {\bfseries 11}
  (2011) 043}, \href{http://arxiv.org/abs/1108.5152}{{\ttfamily arXiv:1108.5152
  [hep-th]}}.

\bibitem{Verlinde:1999fy}
H.~L. Verlinde, ``{Holography and compactification},''
  \href{http://dx.doi.org/10.1016/S0550-3213(00)00224-8}{{\em Nucl. Phys. B}
  {\bfseries 580} (2000) 264--274},
  \href{http://arxiv.org/abs/hep-th/9906182}{{\ttfamily arXiv:hep-th/9906182}}.

\bibitem{Klebanov:2000hb}
I.~R. Klebanov and M.~J. Strassler, ``{Supergravity and a confining gauge
  theory: Duality cascades and chi SB resolution of naked singularities},''
  \href{http://dx.doi.org/10.1088/1126-6708/2000/08/052}{{\em JHEP} {\bfseries
  08} (2000) 052}, \href{http://arxiv.org/abs/hep-th/0007191}{{\ttfamily
  arXiv:hep-th/0007191}}.

\bibitem{Giddings:2001yu}
S.~B. Giddings, S.~Kachru, and J.~Polchinski, ``{Hierarchies from fluxes in
  string compactifications},''
  \href{http://dx.doi.org/10.1103/PhysRevD.66.106006}{{\em Phys. Rev. D}
  {\bfseries 66} (2002) 106006},
  \href{http://arxiv.org/abs/hep-th/0105097}{{\ttfamily arXiv:hep-th/0105097}}.

\bibitem{Karch:2001cw}
A.~Karch and L.~Randall, ``{Localized gravity in string theory},''
  \href{http://dx.doi.org/10.1103/PhysRevLett.87.061601}{{\em Phys. Rev. Lett.}
  {\bfseries 87} (2001) 061601},
  \href{http://arxiv.org/abs/hep-th/0105108}{{\ttfamily arXiv:hep-th/0105108}}.

\bibitem{DHoker:2007hhe}
E.~D'Hoker, J.~Estes, and M.~Gutperle, ``{Exact half-BPS Type IIB interface
  solutions. II. Flux solutions and multi-Janus},''
  \href{http://dx.doi.org/10.1088/1126-6708/2007/06/022}{{\em JHEP} {\bfseries
  06} (2007) 022}, \href{http://arxiv.org/abs/0705.0024}{{\ttfamily
  arXiv:0705.0024 [hep-th]}}.

\bibitem{DHoker:2007zhm}
E.~D'Hoker, J.~Estes, and M.~Gutperle, ``{Exact half-BPS Type IIB interface
  solutions. I. Local solution and supersymmetric Janus},''
  \href{http://dx.doi.org/10.1088/1126-6708/2007/06/021}{{\em JHEP} {\bfseries
  06} (2007) 021}, \href{http://arxiv.org/abs/0705.0022}{{\ttfamily
  arXiv:0705.0022 [hep-th]}}.

\bibitem{Kachru:2003sx}
S.~Kachru, R.~Kallosh, A.~D. Linde, J.~M. Maldacena, L.~P. McAllister, and
  S.~P. Trivedi, ``{Towards inflation in string theory},''
  \href{http://dx.doi.org/10.1088/1475-7516/2003/10/013}{{\em JCAP} {\bfseries
  10} (2003) 013}, \href{http://arxiv.org/abs/hep-th/0308055}{{\ttfamily
  arXiv:hep-th/0308055}}.

\bibitem{Kachru:2019dvo}
S.~Kachru, M.~Kim, L.~Mcallister, and M.~Zimet, ``{de Sitter vacua from ten
  dimensions},'' \href{http://dx.doi.org/10.1007/JHEP12(2021)111}{{\em JHEP}
  {\bfseries 12} (2021) 111}, \href{http://arxiv.org/abs/1908.04788}{{\ttfamily
  arXiv:1908.04788 [hep-th]}}.

\bibitem{Demirtas:2021nlu}
M.~Demirtas, M.~Kim, L.~McAllister, J.~Moritz, and A.~Rios-Tascon, ``{Small
  cosmological constants in string theory},''
  \href{http://dx.doi.org/10.1007/JHEP12(2021)136}{{\em JHEP} {\bfseries 12}
  (2021) 136}, \href{http://arxiv.org/abs/2107.09064}{{\ttfamily
  arXiv:2107.09064 [hep-th]}}.

\bibitem{Randall:2019ent}
L.~Randall, ``{The Boundaries of KKLT},''
  \href{http://dx.doi.org/10.1002/prop.201900105}{{\em Fortsch. Phys.}
  {\bfseries 68} no.~3-4, (2020) 1900105},
  \href{http://arxiv.org/abs/1912.06693}{{\ttfamily arXiv:1912.06693
  [hep-th]}}.

\bibitem{Alishahiha:2004md}
M.~Alishahiha, A.~Karch, E.~Silverstein, and D.~Tong, ``{The dS/dS
  correspondence},'' \href{http://dx.doi.org/10.1063/1.1848341}{{\em AIP Conf.
  Proc.} {\bfseries 743} no.~1, (2004) 393--409},
  \href{http://arxiv.org/abs/hep-th/0407125}{{\ttfamily arXiv:hep-th/0407125}}.

\bibitem{Karch:2003em}
A.~Karch, ``{Autolocalization in de Sitter space},''
  \href{http://dx.doi.org/10.1088/1126-6708/2003/07/050}{{\em JHEP} {\bfseries
  07} (2003) 050}, \href{http://arxiv.org/abs/hep-th/0305192}{{\ttfamily
  arXiv:hep-th/0305192}}.

\bibitem{Dong:2010pm}
X.~Dong, B.~Horn, E.~Silverstein, and G.~Torroba, ``{Micromanaging de Sitter
  holography},'' \href{http://dx.doi.org/10.1088/0264-9381/27/24/245020}{{\em
  Class. Quant. Grav.} {\bfseries 27} (2010) 245020},
  \href{http://arxiv.org/abs/1005.5403}{{\ttfamily arXiv:1005.5403 [hep-th]}}.

\bibitem{Gorbenko:2018oov}
V.~Gorbenko, E.~Silverstein, and G.~Torroba, ``{dS/dS and $ T\overline{T} $},''
  \href{http://dx.doi.org/10.1007/JHEP03(2019)085}{{\em JHEP} {\bfseries 03}
  (2019) 085}, \href{http://arxiv.org/abs/1811.07965}{{\ttfamily
  arXiv:1811.07965 [hep-th]}}.

\bibitem{Coleman:2021nor}
E.~Coleman, E.~A. Mazenc, V.~Shyam, E.~Silverstein, R.~M. Soni, G.~Torroba, and
  S.~Yang, ``{de Sitter Microstates from $T\bar T+\Lambda_2$ and the
  Hawking-Page Transition},'' \href{http://arxiv.org/abs/2110.14670}{{\ttfamily
  arXiv:2110.14670 [hep-th]}}.

\bibitem{Bousso:2022ntt}
R.~Bousso, X.~Dong, N.~Engelhardt, T.~Faulkner, T.~Hartman, S.~H. Shenker, and
  D.~Stanford, ``{Snowmass White Paper: Quantum Aspects of Black Holes and the
  Emergence of Spacetime},'' \href{http://arxiv.org/abs/2201.03096}{{\ttfamily
  arXiv:2201.03096 [hep-th]}}.

\bibitem{Laddha:2020kvp}
A.~Laddha, S.~G. Prabhu, S.~Raju, and P.~Shrivastava, ``{The Holographic Nature
  of Null Infinity},''
  \href{http://dx.doi.org/10.21468/SciPostPhys.10.2.041}{{\em SciPost Phys.}
  {\bfseries 10} no.~2, (2021) 041},
  \href{http://arxiv.org/abs/2002.02448}{{\ttfamily arXiv:2002.02448
  [hep-th]}}.

\bibitem{Raju:2021lwh}
S.~Raju, ``{Failure of the split property in gravity and the information
  paradox},'' \href{http://arxiv.org/abs/2110.05470}{{\ttfamily
  arXiv:2110.05470 [hep-th]}}.

\bibitem{Geng:2021hlu}
H.~Geng, A.~Karch, C.~Perez-Pardavila, S.~Raju, L.~Randall, M.~Riojas, and
  S.~Shashi, ``{Inconsistency of islands in theories with long-range
  gravity},'' \href{http://dx.doi.org/10.1007/JHEP01(2022)182}{{\em JHEP}
  {\bfseries 01} (2022) 182}, \href{http://arxiv.org/abs/2107.03390}{{\ttfamily
  arXiv:2107.03390 [hep-th]}}.

\bibitem{Almheiri:2019psy}
A.~Almheiri, R.~Mahajan, and J.~E. Santos, ``{Entanglement islands in higher
  dimensions},'' \href{http://dx.doi.org/10.21468/SciPostPhys.9.1.001}{{\em
  SciPost Phys.} {\bfseries 9} no.~1, (2020) 001},
  \href{http://arxiv.org/abs/1911.09666}{{\ttfamily arXiv:1911.09666
  [hep-th]}}.

\bibitem{Almheiri:2019hni}
A.~Almheiri, R.~Mahajan, J.~Maldacena, and Y.~Zhao, ``{The Page curve of
  Hawking radiation from semiclassical geometry},''
  \href{http://dx.doi.org/10.1007/JHEP03(2020)149}{{\em JHEP} {\bfseries 03}
  (2020) 149}, \href{http://arxiv.org/abs/1908.10996}{{\ttfamily
  arXiv:1908.10996 [hep-th]}}.

\bibitem{Uhlemann:2021nhu}
C.~F. Uhlemann, ``{Islands and Page curves in 4d from Type IIB},''
  \href{http://dx.doi.org/10.1007/JHEP08(2021)104}{{\em JHEP} {\bfseries 08}
  (2021) 104}, \href{http://arxiv.org/abs/2105.00008}{{\ttfamily
  arXiv:2105.00008 [hep-th]}}.

\bibitem{Rozali:2019day}
M.~Rozali, J.~Sully, M.~Van~Raamsdonk, C.~Waddell, and D.~Wakeham,
  ``{Information radiation in BCFT models of black holes},''
  \href{http://dx.doi.org/10.1007/JHEP05(2020)004}{{\em JHEP} {\bfseries 05}
  (2020) 004}, \href{http://arxiv.org/abs/1910.12836}{{\ttfamily
  arXiv:1910.12836 [hep-th]}}.

\bibitem{Geng:2020qvw}
H.~Geng and A.~Karch, ``{Massive islands},''
  \href{http://dx.doi.org/10.1007/JHEP09(2020)121}{{\em JHEP} {\bfseries 09}
  (2020) 121}, \href{http://arxiv.org/abs/2006.02438}{{\ttfamily
  arXiv:2006.02438 [hep-th]}}.

\bibitem{Porrati:2001db}
M.~Porrati, ``{Higgs phenomenon for 4-D gravity in anti-de Sitter space},''
  \href{http://dx.doi.org/10.1088/1126-6708/2002/04/058}{{\em JHEP} {\bfseries
  04} (2002) 058}, \href{http://arxiv.org/abs/hep-th/0112166}{{\ttfamily
  arXiv:hep-th/0112166}}.

\bibitem{Aharony:2003qf}
O.~Aharony, O.~DeWolfe, D.~Z. Freedman, and A.~Karch, ``{Defect conformal field
  theory and locally localized gravity},''
  \href{http://dx.doi.org/10.1088/1126-6708/2003/07/030}{{\em JHEP} {\bfseries
  07} (2003) 030}, \href{http://arxiv.org/abs/hep-th/0303249}{{\ttfamily
  arXiv:hep-th/0303249}}.

\bibitem{Aharony:2006hz}
O.~Aharony, A.~B. Clark, and A.~Karch, ``{The CFT/AdS correspondence, massive
  gravitons and a connectivity index conjecture},''
  \href{http://dx.doi.org/10.1103/PhysRevD.74.086006}{{\em Phys. Rev. D}
  {\bfseries 74} (2006) 086006},
  \href{http://arxiv.org/abs/hep-th/0608089}{{\ttfamily arXiv:hep-th/0608089}}.

\bibitem{Kiritsis:2006hy}
E.~Kiritsis, ``{Product CFTs, gravitational cloning, massive gravitons and the
  space of gravitational duals},''
  \href{http://dx.doi.org/10.1088/1126-6708/2006/11/049}{{\em JHEP} {\bfseries
  11} (2006) 049}, \href{http://arxiv.org/abs/hep-th/0608088}{{\ttfamily
  arXiv:hep-th/0608088}}.

\bibitem{Geng:2020fxl}
H.~Geng, A.~Karch, C.~Perez-Pardavila, S.~Raju, L.~Randall, M.~Riojas, and
  S.~Shashi, ``{Information Transfer with a Gravitating Bath},''
  \href{http://dx.doi.org/10.21468/SciPostPhys.10.5.103}{{\em SciPost Phys.}
  {\bfseries 10} no.~5, (2021) 103},
  \href{http://arxiv.org/abs/2012.04671}{{\ttfamily arXiv:2012.04671
  [hep-th]}}.

\bibitem{Penington:2019kki}
G.~Penington, S.~H. Shenker, D.~Stanford, and Z.~Yang, ``{Replica wormholes and
  the black hole interior},'' \href{http://arxiv.org/abs/1911.11977}{{\ttfamily
  arXiv:1911.11977 [hep-th]}}.

\bibitem{Chen:2020hmv}
H.~Z. Chen, R.~C. Myers, D.~Neuenfeld, I.~A. Reyes, and J.~Sandor, ``{Quantum
  Extremal Islands Made Easy, Part II: Black Holes on the Brane},''
  \href{http://dx.doi.org/10.1007/JHEP12(2020)025}{{\em JHEP} {\bfseries 12}
  (2020) 025}, \href{http://arxiv.org/abs/2010.00018}{{\ttfamily
  arXiv:2010.00018 [hep-th]}}.

\bibitem{Chen:2020uac}
H.~Z. Chen, R.~C. Myers, D.~Neuenfeld, I.~A. Reyes, and J.~Sandor, ``{Quantum
  Extremal Islands Made Easy, Part I: Entanglement on the Brane},''
  \href{http://dx.doi.org/10.1007/JHEP10(2020)166}{{\em JHEP} {\bfseries 10}
  (2020) 166}, \href{http://arxiv.org/abs/2006.04851}{{\ttfamily
  arXiv:2006.04851 [hep-th]}}.

\bibitem{Geng:2021mic}
H.~Geng, A.~Karch, C.~Perez-Pardavila, S.~Raju, L.~Randall, M.~Riojas, and
  S.~Shashi, ``{Entanglement Phase Structure of a Holographic BCFT in a Black
  Hole Background},'' \href{http://arxiv.org/abs/2112.09132}{{\ttfamily
  arXiv:2112.09132 [hep-th]}}.

\bibitem{Geng:2021wcq}
H.~Geng, Y.~Nomura, and H.-Y. Sun, ``{Information paradox and its resolution in
  de Sitter holography},''
  \href{http://dx.doi.org/10.1103/PhysRevD.103.126004}{{\em Phys. Rev. D}
  {\bfseries 103} no.~12, (2021) 126004},
  \href{http://arxiv.org/abs/2103.07477}{{\ttfamily arXiv:2103.07477
  [hep-th]}}.

\bibitem{Neuenfeld:2021wbl}
D.~Neuenfeld, ``{The Dictionary for Double Holography and Graviton Masses in d
  Dimensions},'' \href{http://arxiv.org/abs/2104.02801}{{\ttfamily
  arXiv:2104.02801 [hep-th]}}.

\bibitem{Omiya:2021olc}
H.~Omiya and Z.~Wei, ``{Causal Structures and Nonlocality in Double
  Holography},'' \href{http://arxiv.org/abs/2107.01219}{{\ttfamily
  arXiv:2107.01219 [hep-th]}}.

\bibitem{Maldacena:2020sxe}
J.~Maldacena and A.~Milekhin, ``{Humanly traversable wormholes},''
  \href{http://dx.doi.org/10.1103/PhysRevD.103.066007}{{\em Phys. Rev. D}
  {\bfseries 103} no.~6, (2021) 066007},
  \href{http://arxiv.org/abs/2008.06618}{{\ttfamily arXiv:2008.06618
  [hep-th]}}.

\bibitem{Mishra:2017zan}
R.~K. Mishra and R.~Sundrum, ``{Asymptotic Symmetries, Holography and
  Topological Hair},'' \href{http://dx.doi.org/10.1007/JHEP01(2018)014}{{\em
  JHEP} {\bfseries 01} (2018) 014},
  \href{http://arxiv.org/abs/1706.09080}{{\ttfamily arXiv:1706.09080
  [hep-th]}}.

\end{thebibliography}\endgroup
 
\end{document}